\documentclass[aps,prl,twocolumn,amsmath,amsfonts,floatfix,groupaddress]{revtex4-1}
\usepackage{natbib}
\usepackage{graphicx,amssymb}
\usepackage{color}
\newcommand{\rev}[1]{{\color{black}{#1}}}
\newcommand{\revv}[1]{{\color{black}{#1}}}

\begin{document}

\title{Self-induced heterogeneity in deeply supercooled liquids}

\author{Ludovic Berthier}

\affiliation{Laboratoire Charles Coulomb (L2C), Universit\'e de Montpellier, CNRS, 34095 Montpellier, France}

\affiliation{Department of Chemistry, University of Cambridge, Lensfield Road, Cambridge CB2 1EW, United Kingdom}

\begin{abstract}
A theoretical treatment of deeply supercooled liquids is difficult because their properties emerge from spatial inhomogeneities that are self-induced, transient, and nanoscopic. I use computer simulations to analyse self-induced static and dynamic heterogeneity in equilibrium systems approaching the experimental glass transition. I characterise the broad sample-to-sample fluctuations of salient dynamic and thermodynamic properties in elementary mesoscopic systems. Findings regarding local lifetimes and distributions of dynamic heterogeneity are in excellent agreement with recent single molecule studies. Surprisingly broad thermodynamic fluctuations are also found, which correlate well with dynamics fluctuations, thus providing a local test of the thermodynamic origin of slow dynamics.
\end{abstract}

\date{\today}

\maketitle

The physical properties of crystals follow from the periodic repetition of a unit cell~\cite{Ashcroft}. By contrast, amorphous solids are aperiodic structures displaying frozen spatial inhomogeneities~\cite{angell1995formation,ediger1996supercooled,wolynes2012structural} which require special theoretical approaches~\cite{singh1985hard,parisi2020theory,berthier2011theoretical}. Simple fluids are also structurally disordered, but spatial fluctuations are weak enough that physical properties can be accurately predicted without explicitely dealing with disorder~\cite{hansen1990theory}. Deeply supercooled liquids represent a conceptual challenge as they are structurally disordered, with physical properties driven by the existence of spatial heterogeneities, but these fluctuations are transient, induced by the competition between frustrated particle interactions and thermal fluctuations, and of modest spatial extension~\cite{berthier2011theoretical}. 

\begin{figure*}
\begin{center}
\includegraphics[width=\textwidth]{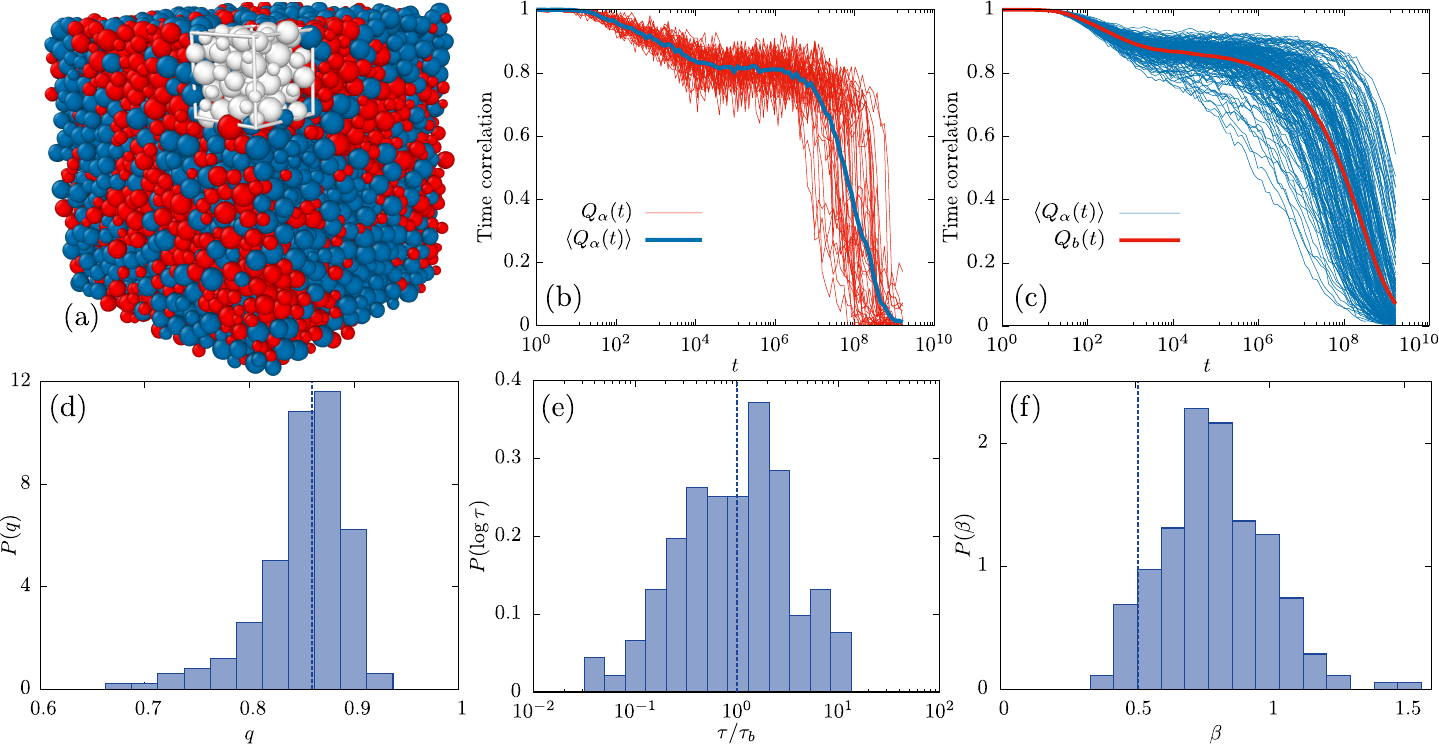}
\end{center}
\caption{
(a) I highlight a subsystem comprising $N=111$ particles in a larger system of $N=8000$ hard spheres, whose spatially heterogeneous dynamics is illustrated by distinguishing fast (red) and slow (blue) particles near $\tau_b$ at $P=30$.
(b) Time decay of the overlap $Q_\alpha(t)$ in 40 isoconfigurational trajectories, and the resulting average $\langle Q_\alpha(t) \rangle$ for a single configuration $\alpha$ with $N=111$ and $P=30$.   
(c) Time decay of isoconfigurational average $\langle Q_\alpha(t) \rangle$ for 200 independent configurations with $N=111$ and $P=30$ and resulting bulk average $Q_b(t)$.  
(d,e,f) Sample-to-sample fluctuations of the plateau height $q$, characteristic timescale $\tau$ and stretching exponent $\beta$ for $P=30$. Bulk values are marked with vertical dashed lines.} 
\label{fig:local}
\end{figure*}

The physics of bulk supercooled liquids has received considerable attention~\cite{angell1995formation,singh1985hard,parisi2020theory,ediger1996supercooled,berthier2011theoretical,wolynes2012structural}. It is established that dynamics becomes spatially heterogeneous over a characteristic lengthscale that grows modestly approaching the experimental glass transition~\cite{ediger2000spatially,richert2002heterogeneous,berthier2011dynamical}. This mesoscopic dynamic heterogeneity is illustrated in Fig.~\ref{fig:local}(a), obtained from simulating a hard sphere model introduced shortly. The microstructure of supercooled liquids is very complex and fluctuates broadly at the particle scale~\cite{royall2015role}. Structural heterogeneity reflects the large variety of amorphous packings that particles can locally adopt. These multiple disordered structures are the real-space signature of a rugged free energy landscape, but \rev{this local disorder is {\it self-induced}}, as the original Hamiltonian does not contain quenched random interactions. 

Bulk physical properties in supercooled liquids emerge as an ensemble average over locally distributed quantities~\cite{richert2002heterogeneous}. Ideally, one would like to understand the fluctuations of local thermodynamic and structural properties to then infer heterogeneous relaxation processes~\cite{xia2001microscopic}. The disorder strength also represents the microscopic mechanism by which the ideal glass transition may disappear in finite dimensions~\cite{doi:10.1063/1.3009827,PhysRevLett.112.175701,dzero2009replica}, but \rev{this was never measured directly.} The emblematic stretched exponential decay of time correlations~\cite{doi:10.1063/1.466117,cardona2007history} reflects a  distribution of local relaxation functions with fluctuating shapes and timescales~\cite{ediger2000spatially,richert2002heterogeneous}. Recent single molecule~\cite{zondervan2007local,paeng2015ideal,manz2018single,manz2019correlating} and electron correlation microscopy~\cite{zhang2018spatially} studies give direct access to such distributions, but the experimental evidence of fluctuating local distributions \rev{is contradicted simulations}~\cite{shang2019local}. The structural origin of dynamic heterogeneity remains a wide open question. Previous work discussed either particle-based structural indicators~\cite{widmer2008irreversible,royall2015role,schoenholz2016structural,tong2018revealing} or non-causal correlations between bulk structural and dynamical quantities~\cite{richert1998dynamics,sastry2001relationship,larini2008universal,gundermann2014dynamic,ozawa2019does}.
Statically, self-induced disorder is at the core of the mean-field theory of glasses~\cite{parisi2020theory} and random first order transition theory~\cite{kirkpatrick1989scaling}, and, ultimately, the reason why configurational entropy is central to glass studies~\cite{berthier2019configurational}. Some consequences of self-induced disorder on bulk thermodynamics have been analysed before in simulations~\cite{berthier2015evidence,coslovich2016structure,guiselin2020random}, but \rev{very little is known about the distribution of these fluctuations in realistic models and their temperature evolution. Most importantly, a link between local entropy fluctuations and structural relaxation has never been established, although this idea, proposed more than 50 years ago~\cite{adam1965temperature}, has been analysed in countless publications.} 

\rev{I numerically analyse self-induced heterogeneity by studying dynamic and static properties of statistically independent mesoscopic systems, attacking three different lines of investigations.} As pioneered by Heuer and coworkers~\cite{buchner1999potential,la2006relation,heuer2008exploring} and illustrated in Fig.~\ref{fig:local}(a), the study of small systems grants access to fluctuations created by the specific packing adopted locally in each configuration. Also key to our study is the use of the swap Monte Carlo algorithm~\cite{ninarello2017models} which provides fast equilibration down to the experimental glass transition and \rev{allows to probe, for the first time, local fluctuations in the appropriate temperature regime.} \rev{We first characterise the dynamics} of elementary systems using isoconfigurational averaging~\cite{widmer2004reproducible} to obtain smooth time correlations for each sample. We measure both sample-to-sample fluctuations of the relaxation dynamics and the local lifetime of the dynamic heterogeneity, which compare well with recent experiments~\cite{paeng2015ideal,manz2018single,manz2019correlating,zhang2018spatially}. \rev{Second, we characterise} sample-to-sample fluctuations of the Franz-Parisi free energy~\cite{franz1997phase}, which is directly related, in bulk systems, to the configurational entropy~\cite{berthier2014novel}. \rev{Third, we show that} thermodynamic and dynamic fluctuations are correlated, which amounts to validating \revv{a local version of the Adam-Gibbs relation.} 

We simulate three-dimensional non-additive polydisperse hard spheres with a flat distribution of particle diameters and polydispersity around 23\% (details as in \cite{berthier2020measure}). To probe elementary systems, we use the smallest possible system size with periodic boundary conditions. Too small systems have an incorrect pair correlation~\cite{buchner1999potential} and can be structurally unstable~\cite{brumer2004numerical}. We settle for $N=111$ particles (a smaller system with $N=71$ too easily crystallised), which is smaller than the extension of the dynamic heterogeneity, see Fig.~\ref{fig:local}(a), \rev{but has an average behaviour comparable to larger systems.} We perform constant pressure Monte Carlo simulations at constant temperature \revv{(set to $T=1$, for convenience in units where the Boltzmann constant $k_B$ is also unity)}, and use the pressure $P$ as control parameter \rev{(constant density simulations would yield equivalent results).} {This is also convenient} as $P$ \revv{(or, rather the compressibility factor $P/(\rho k_B T)$, where $\rho$ is the number density)} plays for hard spheres a role strictly equivalent to $1/T$ in systems with soft interactions~\cite{berthier2009glass}. For this system~\cite{berthier2020measure}, $P=20$ corresponds to the onset of slow dynamics, $P=26$ to the mode-coupling crossover, and $P = 37$ is a conservative estimate of the experimental glass transition. One Monte Carlo timestep represents $N$ attempts to perform a translational move. The unit length is the average particle diameter, and \rev{the particle sizes are kept constant in each sample.}

We analyse sample-to-sample fluctuations of the structural relaxation up to $P=30$, much beyond the mode-coupling crossover. We run 40 distinct trajectories starting from each configuration $\alpha$ equilibrated using swap Monte Carlo during which we record the time overlap $Q_\alpha(t) = 1/N \sum_i \theta(a - |{\bf r}_i^\alpha(t)-{\bf r}_i^\alpha(0)|)$, with $a=0.2$, ${\bf r}_i^\alpha(t)$ the position of particle $i$ at time $t$ starting from configuration $\alpha$ at $t=0$, and $\theta(x)$ the Heaviside function \rev{(this choice ensures that the overlap probes equivalent physics to the self-intermediate scattering function~\cite{berthier2017configurational,ozawa2017does}).} 
By averaging over distinct trajectories from the same initial condition $\alpha$ we obtain the isoconfigurational average $\langle Q_\alpha (t) \rangle$, see Fig.~\ref{fig:local}(b). Whereas $Q_\alpha(t)$ is not smooth and fluctuates among \rev{individual trajectories}, $\langle Q_\alpha(t) \rangle$ has a smooth time dependence which fully characterizes the relaxation dynamics of configuration $\alpha$. The isoconfigurational average advantageously replaces the time average needed in single molecule studies. In a final step, we perform an additional average over $200$ independent configurations to obtain the bulk correlation function, $Q_b(t) = \overline{\langle Q_\alpha(t) \rangle}$, see Fig.~\ref{fig:local}(c). To analyse the fluctuations, each isoconfigurational correlation is fitted to a stretched exponential, 
\begin{equation}
\langle Q_\alpha(t) \rangle = q \exp \left( - (t/\tau)^\beta \right),
\end{equation} 
where $(q, \beta, \tau)$ are three fitting parameters, taking values $(q_b,\beta_b,\tau_b)$ for the bulk function $Q_b(t)$. 

In Figs.~\ref{fig:local}(d-f) we show histograms of their sample-to-sample fluctuations for $P=30$. The plateau height $q$ has surprisingly large fluctuations, of order $20\%$, with a pronounced tail towards small values. The relaxation time $\tau$ varies over about 3 orders of magnitude, with again a significant tail corresponding to systems relaxing much faster than the bulk. This broad distribution represents a direct probe of spatially heterogeneous dynamics. Surprisingly, the stretching exponent $\beta$ is also broadly distributed, with a most probable value $\overline{\beta} \simeq 0.75$ much larger than the bulk $\beta_b \simeq 0.5$. Elementary systems do not have the same stretching exponent as the bulk, and sample-to-sample fluctuations are massive. \rev{As shown in Fig.~\ref{fig:local}(b), the overlap in individual trajectories $Q_\alpha(t)$ is typically more compressed than its isoconfigurational average, $\langle Q_\alpha(t) \rangle$, but are quite stretched in samples with small $\beta$.} \rev{Our results contradict a recent study obtained using frozen cavities~\cite{shang2019local}, perhaps because relatively large cavities were used.} Our simulations reveal that the streched exponential relaxation of the bulk system stems both from a broad distribution of local relaxation times and from strong deviations from a local exponential decay. They also support results from single molecule studies~\cite{paeng2015ideal} and electron correlation microscopy~\cite{zhang2018spatially}, and differ from \rev{some} earlier conclusions~\cite{bohmer1998nature}.

\begin{figure}
\includegraphics[width=8.5cm]{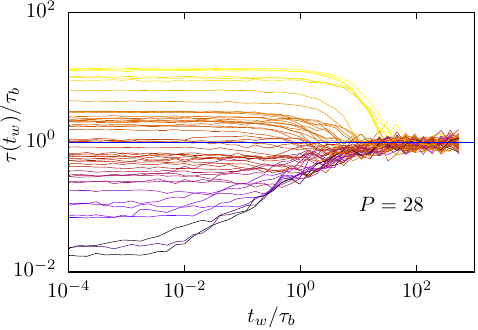}
\caption{Evolution of the local relaxation time $\tau(t_w)$ as the system progressively loses the memory of its initial condition and returns to the average behaviour. Fast and slow samples display asymmetric behaviour, but have a similar lifetime of about $50 \tau_b$.}
\label{fig:lifetime}
\end{figure}

Our strategy probes structural relaxation locally without any time averaging. We can then ask how long the dynamic heterogeneity persists, i.e. how long it takes a spontaneous fluctuation to return to average behaviour. To this end, we promote the overlap to a two-time correlation, $Q_\alpha(t_w+t,t_w)$, by recording the dynamics between times $t_w$ and $t_w+t$ starting from configuration $\alpha$ at $t_w=0$. The case $t_w=0$ was considered in Fig.~\ref{fig:local}. In the opposite limit, $t_w \to \infty$, the system loses memory of the initial condition: 
$\lim_{t_w \to \infty} \langle Q_\alpha(t_w, t_w+t) \rangle = Q_b(t)$.
Our goal is to quantify how this limit is reached in different samples, allowing us to define a \rev{local lifetime for dynamic heterogeneity, as opposed to global ones discussed before~\cite{paeng2015ideal}.} We define $\tau(t_w)$ from the time decay of $\langle Q_\alpha(t_w+t,t_w) \rangle$ for each configuration $\alpha$.

The results are shown in Fig.~\ref{fig:lifetime} for $P=28$. For each sample, we observe that $\tau(t_w)$ displays a plateau of duration $t_w \sim \tau(t_w=0)$, which simply confirms that no useful dynamics is happening at times shorter than $\tau$. However, all samples (fast and slow) return to the bulk value $\tau_b$ over a similar timescale $t_w \approx 50 \tau_b$, but they do in a very asymmetric manner reminiscent of nonlinear aging studies~\cite{angell2000relaxation}. Over the explored time window, no pronounced pressure dependence was detected for this lifetime. These results are again in harmony with experimental findings from single molecule studies~\cite{paeng2015ideal} and \rev{earlier determinations of heterogeneity lifetime~\cite{kuebler1997glass}.}   

\begin{figure*}
\includegraphics[width=\textwidth]{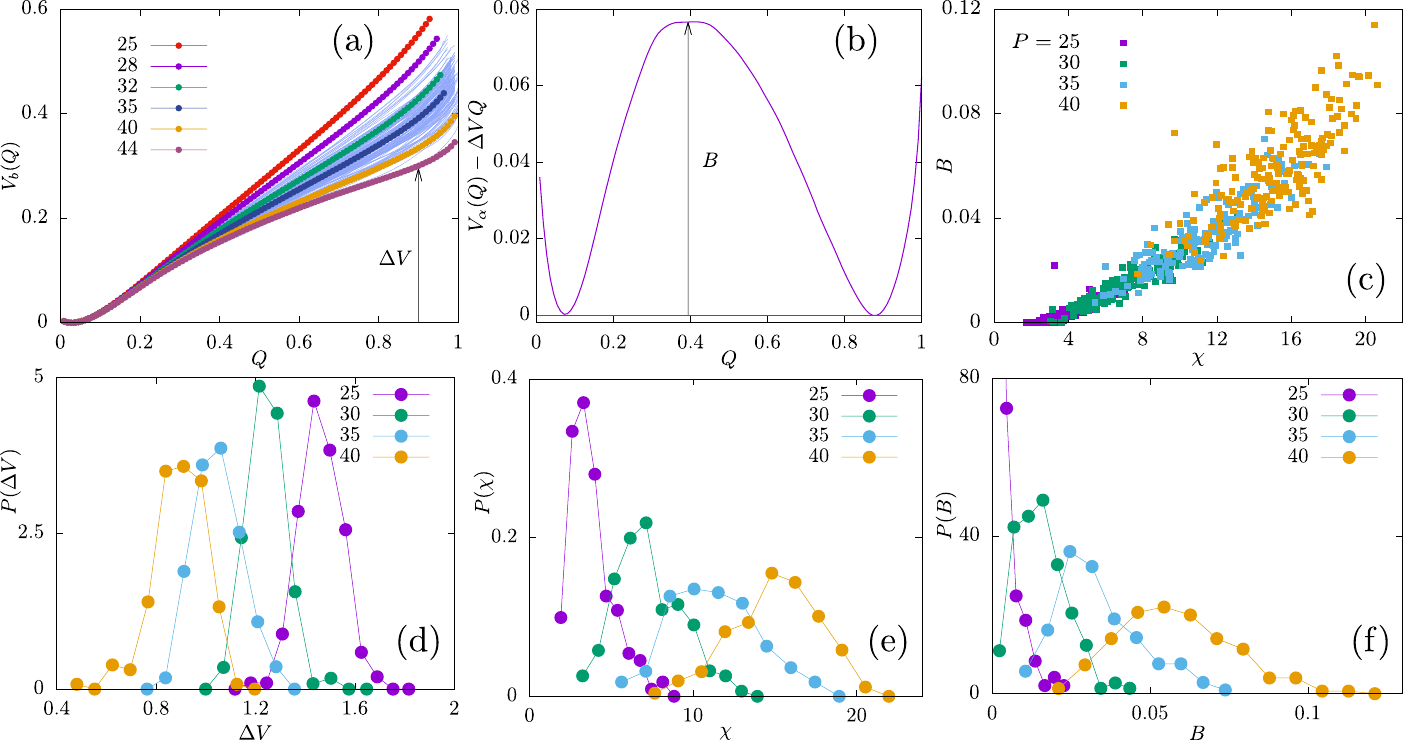}
\caption{(a) Evolution of the bulk Franz-Parisi potential $V_b(Q)$ with pressure. For $P=35$, we also show $V_\alpha(Q)$ from multiple independent samples (thin lines), showing the breadth of self-induced free energy fluctuations. 
(b) Definition of the free energy difference $\Delta V$ and the barrier $B$ for an individual sample at $P=35$. 
(c) Scatter plot of the the barrier $B$ versus the susceptibility $\chi$ for independent samples at different pressures. 
(d,e,f) Histogram of sample-to-sample fluctuations of $\Delta V$, $\chi$ and $B$.}
\label{fig:FP2}
\end{figure*}

\rev{We turn to thermodynamic fluctuations.} The order parameter for a static glass transition is the overlap $Q_{\alpha \beta}=1/N \sum_{i,j} \theta(a-|{\bf r}_i^\alpha - {\bf r}_j^\beta|)$ between a pair $(\alpha, \beta)$ of equilibrium configurations. From the fluctuations of the overlap, the bulk Franz-Parisi (FP) free energy $V_b(Q)$ is constructed~\cite{franz1997phase}. 
This is well studied numerically~\cite{berthier2013overlap,berthier2014novel,guiselin2020random}. Here, we focus instead on sample-to-sample fluctuations, \rev{about which little is known at low temperatures.} The equilibrium fluctuations of $Q_{\alpha \beta}$ between a fixed reference configuration $\alpha$ and fluctuating configurations $\beta$ define the distribution $P_\alpha(Q) = \langle \delta (Q- Q_{\alpha \beta}) \rangle$, which yields $V_\alpha(Q) = -(T/N) \log P_\alpha(Q)$ for each sample $\alpha$. The bulk free energy $V_b(Q)$ is obtained after averaging over $\alpha$: $V_b(Q) = \overline{V_\alpha(Q)}$.  
 
Fig.~\ref{fig:FP2}(a) shows the evolution with $P$ of the bulk FP potential $V_b(Q)$. For $P=35$ (near the experimental glass transition), we show $V_\alpha(Q)$ for individual samples. The evolution of $V_b(Q)$ resembles previous results~\cite{berthier2013overlap}, and demonstrates the approach to a random first-order phase transition (possibly) occurring at larger $P$. The non-convexity of $V_b(Q)$ at large $P$ stems from using a small $N$ and is absent in larger systems~\cite{guiselin2020random}. An estimate of the bulk configurational entropy~\cite{berthier2014novel} follows the free energy difference $\Delta V_b$ between high and low $Q$ values, see Fig.~\ref{fig:FP2}(a).   

The evolution of $V_b(Q)$ is smooth, but \rev{the sample-to-sample fluctuations observed at each state point are surprisingly large.} For $P=35$, we obtain a range of $V_\alpha(Q)$ resembling the measured $V_b(Q)$ over the range $P \in [28, 40]$, i.e. from the mode-coupling crossover to below the experimental glass transition. \rev{This shows that spatial fluctuations of the configurational entropy are surprisingly large, which, to our knowledge, has never been observed before.}

\rev{To start quantifying these observations, we introduce three measures: $(\Delta V, B, \chi)$}. For each sample, $\Delta V$ is the critical value of the field needed for the function $V_\alpha(Q) - \Delta V Q$ to have two minima of the same height [see Fig.~\ref{fig:FP2}(b)], and thus to induce a discontinuous transition between high and low $Q$ phases; $B$ is the barrier separating the two minima, see Fig.~\ref{fig:FP2}(b); the susceptibility $\chi$ is the variance of overlap fluctuations at coexistence. Scatter plots reveal that these three quantities are strongly correlated [\rev{for instance} $B$ vs. $\chi$ in Fig.~\ref{fig:FP2}(c)], and \rev{thus they describe well local fluctuations} of the FP free energy. The histograms in Figs.~\ref{fig:FP2}(d-f) show the evolution of these fluctuations over a broad range of pressures. \rev{There is considerable overlap between the distributions measured at well separated state points, reflecting the large amount of disorder in supercooled liquids.} These fluctuations represent a coarse-grained, agnostic, non-mechanical measure of the thermodynamic stability of the local particle packing, with no reference to a potential energy landscape~\cite{la2006relation} (which is not appropriate for hard spheres anyway). These results also \rev{confirm and quantify} the central role played by self-induced disorder in the thermodynamics of supercooled liquids, thus reinforcing analogies with random field Ising models~\cite{biroli2014random,doi:10.1063/1.3009827,berthier2015evidence,biroli2018random,guiselin2020random}. 

\rev{We come to the third point of the paper. Both the bulk relaxation time $\tau_b$ and FP potential $V_b(Q)$ (therefore, the configurational entropy $\approx \Delta V_b$) evolve noticeably with $P$. A ``correlation'' necessarily} relates $\log \tau_b$ to $\Delta V_b$, but this does not \rev{imply any causality between configurational entropy and relaxation time.} All previous work thus instead tested the validity and universality of the Adam-Gibbs relation~\cite{richert1998dynamics,sastry2001relationship,ozawa2019does}, \revv{which reads $\log (\tau_b /\tau_\infty) =  \frac{P}{\rho k_B T} \frac{k_B T}{\Delta V_b} \sim P/\Delta V_b$, using notations where physical dimensions are transparent.} Collecting static and dynamic sample-to-sample fluctuations, we can directly test whether local fluctuations of the configurational entropy correlate with local fluctuations of the dynamics. \rev{Such demanding test of a putative structure-dynamics correlation has never been performed.} Our approach also \rev{differs from most parallel studies~\cite{widmer2008irreversible,royall2015role,schoenholz2016structural,tong2018revealing},} because we (intentionally~\cite{berthier2007structure}) do not work at the single particle level.

\begin{figure}
\includegraphics[width=8.5cm]{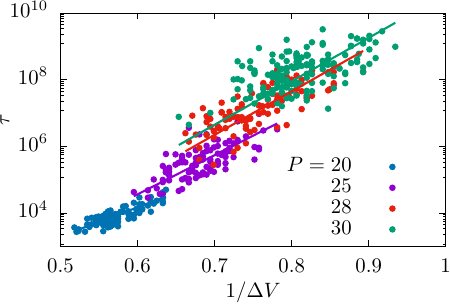}
\caption{Scatter plot of the relaxation time $\tau$ of individual samples against the free energy difference $\Delta V$. The straight lines represent a local version of the Adam-Gibbs relation where $\log \tau \propto (P/\Delta V)$. This shows that self-induced dynamic and thermodynamic fluctuations are locally correlated.}
\label{fig:ag}
\end{figure}

A scatter plot of $\log \tau$ vs. $1/\Delta V$ independently obtained for multiple samples at multiple equilibrium state points is shown in Fig.~\ref{fig:ag} to test \revv{the Adam-Gibbs relation} directly at the level of local fluctuations. \rev{Note that each point represents a substantial numerical treatment to obtain accurate and independent values for $\tau$ (using isoconfigurational dynamic average) and $\Delta V$ (using umbrella sampling techniques). The observed correlation is quite good and the data align along the Adam-Gibbs relation, shown with the straight lines. Interestingly, since $B \sim 1/\Delta V$, this correlation can can be recast as $\log \tau \sim B$, suggesting that $B$ represents a meaningful candidate for the effective barrier to structural relaxation.} By contrast, we find that the plateau height $q$ (local Debye Waller factor) correlates weakly with the dynamics, which appears to contradict models where short-time dynamics is used to infer structural relaxation~\cite{larini2008universal,gundermann2014dynamic}. \rev{Several more tests along} these lines could be performed following the methods introduced above, \rev{and the quality of the various correlations could be quantified further.} 

Deeply supercooled liquids display broad distributions of \rev{their local physical properties,} which are \rev{extensively probed here numerically at extremely low temperatures} by analysing mesoscopic systems \rev{equilibrated with a swap Monte Carlo algorithm}. Each configuration \rev{is ``typical'', yet each sample displays static and dynamic properties} that fluctuate wildly, directly impacting \rev{all observed bulk properties.} Our results \rev{directly establish} that deeply supercooled liquids are found, at the mesoscale, in a very large number of distinct packings with \rev{distinct properties.} This {\it self-induced heterogeneity} represents the real space signature of the metaphorical rugged free energy landscape. Static fluctuations are central to random first order transition theory, but they also exist in simpler plaquette models~\cite{jack2016phase}, \rev{although} both approaches provide distinct general perspectives on their dynamical consequences~\cite{biroli2012random,jack2005caging}.
We focused \rev{here} on an important subset of questions regarding dynamic heterogeneity and thermodynamic fluctuations, but many more can be tackled using similar tools, from experimentally-relevant questions (e.g., rheology) to fundamental ones (e.g., extracting coupling constants of effective field theories). 
  
\begin{acknowledgments}
I thank M. Ediger, L. Kaufman, D. Reichman and C. Scalliet for encouragments and useful exchanges. I borrowed the title from J.-P. Bouchaud. My work is supported by a grant from the Simons Foundation (Grant No. 454933).
\end{acknowledgments}

\bibliography{glasses.bib}{}

\begin{thebibliography}{60}%
\makeatletter
\providecommand \@ifxundefined [1]{%
 \@ifx{#1\undefined}
}%
\providecommand \@ifnum [1]{%
 \ifnum #1\expandafter \@firstoftwo
 \else \expandafter \@secondoftwo
 \fi
}%
\providecommand \@ifx [1]{%
 \ifx #1\expandafter \@firstoftwo
 \else \expandafter \@secondoftwo
 \fi
}%
\providecommand \natexlab [1]{#1}%
\providecommand \enquote  [1]{``#1''}%
\providecommand \bibnamefont  [1]{#1}%
\providecommand \bibfnamefont [1]{#1}%
\providecommand \citenamefont [1]{#1}%
\providecommand \href@noop [0]{\@secondoftwo}%
\providecommand \href [0]{\begingroup \@sanitize@url \@href}%
\providecommand \@href[1]{\@@startlink{#1}\@@href}%
\providecommand \@@href[1]{\endgroup#1\@@endlink}%
\providecommand \@sanitize@url [0]{\catcode `\\12\catcode `\$12\catcode
  `\&12\catcode `\#12\catcode `\^12\catcode `\_12\catcode `\%12\relax}%
\providecommand \@@startlink[1]{}%
\providecommand \@@endlink[0]{}%
\providecommand \url  [0]{\begingroup\@sanitize@url \@url }%
\providecommand \@url [1]{\endgroup\@href {#1}{\urlprefix }}%
\providecommand \urlprefix  [0]{URL }%
\providecommand \Eprint [0]{\href }%
\providecommand \doibase [0]{http://dx.doi.org/}%
\providecommand \selectlanguage [0]{\@gobble}%
\providecommand \bibinfo  [0]{\@secondoftwo}%
\providecommand \bibfield  [0]{\@secondoftwo}%
\providecommand \translation [1]{[#1]}%
\providecommand \BibitemOpen [0]{}%
\providecommand \bibitemStop [0]{}%
\providecommand \bibitemNoStop [0]{.\EOS\space}%
\providecommand \EOS [0]{\spacefactor3000\relax}%
\providecommand \BibitemShut  [1]{\csname bibitem#1\endcsname}%
\let\auto@bib@innerbib\@empty
\bibitem [{\citenamefont {Ashcroft}\ and\ \citenamefont
  {Mermin}(1976)}]{Ashcroft}%
  \BibitemOpen
  \bibfield  {author} {\bibinfo {author} {\bibfnamefont {N.}~\bibnamefont
  {Ashcroft}}\ and\ \bibinfo {author} {\bibfnamefont {N.}~\bibnamefont
  {Mermin}},\ }\href@noop {} {\emph {\bibinfo {title} {{Solid State
  Physics}}}}\ (\bibinfo  {publisher} {Saunders College},\ \bibinfo {address}
  {Philadelphia},\ \bibinfo {year} {1976})\BibitemShut {NoStop}%
\bibitem [{\citenamefont {Angell}(1995)}]{angell1995formation}%
  \BibitemOpen
  \bibfield  {author} {\bibinfo {author} {\bibfnamefont {C.~A.}\ \bibnamefont
  {Angell}},\ }\href@noop {} {\bibfield  {journal} {\bibinfo  {journal}
  {Science}\ }\textbf {\bibinfo {volume} {267}},\ \bibinfo {pages} {1924}
  (\bibinfo {year} {1995})}\BibitemShut {NoStop}%
\bibitem [{\citenamefont {Ediger}\ \emph {et~al.}(1996)\citenamefont {Ediger},
  \citenamefont {Angell},\ and\ \citenamefont {Nagel}}]{ediger1996supercooled}%
  \BibitemOpen
  \bibfield  {author} {\bibinfo {author} {\bibfnamefont {M.~D.}\ \bibnamefont
  {Ediger}}, \bibinfo {author} {\bibfnamefont {C.~A.}\ \bibnamefont {Angell}},
  \ and\ \bibinfo {author} {\bibfnamefont {S.~R.}\ \bibnamefont {Nagel}},\
  }\href@noop {} {\bibfield  {journal} {\bibinfo  {journal} {The journal of
  physical chemistry}\ }\textbf {\bibinfo {volume} {100}},\ \bibinfo {pages}
  {13200} (\bibinfo {year} {1996})}\BibitemShut {NoStop}%
\bibitem [{\citenamefont {Wolynes}\ and\ \citenamefont
  {Lubchenko}(2012)}]{wolynes2012structural}%
  \BibitemOpen
  \bibfield  {author} {\bibinfo {author} {\bibfnamefont {P.~G.}\ \bibnamefont
  {Wolynes}}\ and\ \bibinfo {author} {\bibfnamefont {V.}~\bibnamefont
  {Lubchenko}},\ }\href@noop {} {\emph {\bibinfo {title} {Structural glasses
  and supercooled liquids: Theory, experiment, and applications}}}\ (\bibinfo
  {publisher} {John Wiley \& Sons},\ \bibinfo {year} {2012})\BibitemShut
  {NoStop}%
\bibitem [{\citenamefont {Singh}\ \emph {et~al.}(1985)\citenamefont {Singh},
  \citenamefont {Stoessel},\ and\ \citenamefont {Wolynes}}]{singh1985hard}%
  \BibitemOpen
  \bibfield  {author} {\bibinfo {author} {\bibfnamefont {Y.}~\bibnamefont
  {Singh}}, \bibinfo {author} {\bibfnamefont {J.}~\bibnamefont {Stoessel}}, \
  and\ \bibinfo {author} {\bibfnamefont {P.}~\bibnamefont {Wolynes}},\
  }\href@noop {} {\bibfield  {journal} {\bibinfo  {journal} {Physical review
  letters}\ }\textbf {\bibinfo {volume} {54}},\ \bibinfo {pages} {1059}
  (\bibinfo {year} {1985})}\BibitemShut {NoStop}%
\bibitem [{\citenamefont {Parisi}\ \emph {et~al.}(2020)\citenamefont {Parisi},
  \citenamefont {Urbani},\ and\ \citenamefont {Zamponi}}]{parisi2020theory}%
  \BibitemOpen
  \bibfield  {author} {\bibinfo {author} {\bibfnamefont {G.}~\bibnamefont
  {Parisi}}, \bibinfo {author} {\bibfnamefont {P.}~\bibnamefont {Urbani}}, \
  and\ \bibinfo {author} {\bibfnamefont {F.}~\bibnamefont {Zamponi}},\
  }\href@noop {} {\emph {\bibinfo {title} {Theory of simple glasses: exact
  solutions in infinite dimensions}}}\ (\bibinfo  {publisher} {Cambridge
  University Press},\ \bibinfo {year} {2020})\BibitemShut {NoStop}%
\bibitem [{\citenamefont {Berthier}\ and\ \citenamefont
  {Biroli}(2011)}]{berthier2011theoretical}%
  \BibitemOpen
  \bibfield  {author} {\bibinfo {author} {\bibfnamefont {L.}~\bibnamefont
  {Berthier}}\ and\ \bibinfo {author} {\bibfnamefont {G.}~\bibnamefont
  {Biroli}},\ }\href@noop {} {\bibfield  {journal} {\bibinfo  {journal} {Rev.
  Mod. Phys}\ }\textbf {\bibinfo {volume} {83}},\ \bibinfo {pages} {587}
  (\bibinfo {year} {2011})}\BibitemShut {NoStop}%
\bibitem [{\citenamefont {Hansen}\ and\ \citenamefont
  {McDonald}(1990)}]{hansen1990theory}%
  \BibitemOpen
  \bibfield  {author} {\bibinfo {author} {\bibfnamefont {J.-P.}\ \bibnamefont
  {Hansen}}\ and\ \bibinfo {author} {\bibfnamefont {I.~R.}\ \bibnamefont
  {McDonald}},\ }\href@noop {} {\emph {\bibinfo {title} {Theory of simple
  liquids}}}\ (\bibinfo  {publisher} {Elsevier},\ \bibinfo {year}
  {1990})\BibitemShut {NoStop}%
\bibitem [{\citenamefont {Ediger}(2000)}]{ediger2000spatially}%
  \BibitemOpen
  \bibfield  {author} {\bibinfo {author} {\bibfnamefont {M.~D.}\ \bibnamefont
  {Ediger}},\ }\href@noop {} {\bibfield  {journal} {\bibinfo  {journal} {Annual
  review of physical chemistry}\ }\textbf {\bibinfo {volume} {51}},\ \bibinfo
  {pages} {99} (\bibinfo {year} {2000})}\BibitemShut {NoStop}%
\bibitem [{\citenamefont {Richert}(2002)}]{richert2002heterogeneous}%
  \BibitemOpen
  \bibfield  {author} {\bibinfo {author} {\bibfnamefont {R.}~\bibnamefont
  {Richert}},\ }\href@noop {} {\bibfield  {journal} {\bibinfo  {journal}
  {Journal of Physics: Condensed Matter}\ }\textbf {\bibinfo {volume} {14}},\
  \bibinfo {pages} {R703} (\bibinfo {year} {2002})}\BibitemShut {NoStop}%
\bibitem [{\citenamefont {Berthier}\ \emph {et~al.}(2011)\citenamefont
  {Berthier}, \citenamefont {Biroli}, \citenamefont {Bouchaud}, \citenamefont
  {Cipelletti},\ and\ \citenamefont {van Saarloos}}]{berthier2011dynamical}%
  \BibitemOpen
  \bibfield  {author} {\bibinfo {author} {\bibfnamefont {L.}~\bibnamefont
  {Berthier}}, \bibinfo {author} {\bibfnamefont {G.}~\bibnamefont {Biroli}},
  \bibinfo {author} {\bibfnamefont {J.-P.}\ \bibnamefont {Bouchaud}}, \bibinfo
  {author} {\bibfnamefont {L.}~\bibnamefont {Cipelletti}}, \ and\ \bibinfo
  {author} {\bibfnamefont {W.}~\bibnamefont {van Saarloos}},\ }\href@noop {}
  {\emph {\bibinfo {title} {Dynamical heterogeneities in glasses, colloids, and
  granular media}}},\ Vol.\ \bibinfo {volume} {150}\ (\bibinfo  {publisher}
  {OUP Oxford},\ \bibinfo {year} {2011})\BibitemShut {NoStop}%
\bibitem [{\citenamefont {Royall}\ and\ \citenamefont
  {Williams}(2015)}]{royall2015role}%
  \BibitemOpen
  \bibfield  {author} {\bibinfo {author} {\bibfnamefont {C.~P.}\ \bibnamefont
  {Royall}}\ and\ \bibinfo {author} {\bibfnamefont {S.~R.}\ \bibnamefont
  {Williams}},\ }\href@noop {} {\bibfield  {journal} {\bibinfo  {journal}
  {Physics Reports}\ }\textbf {\bibinfo {volume} {560}},\ \bibinfo {pages} {1}
  (\bibinfo {year} {2015})}\BibitemShut {NoStop}%
\bibitem [{\citenamefont {Xia}\ and\ \citenamefont
  {Wolynes}(2001)}]{xia2001microscopic}%
  \BibitemOpen
  \bibfield  {author} {\bibinfo {author} {\bibfnamefont {X.}~\bibnamefont
  {Xia}}\ and\ \bibinfo {author} {\bibfnamefont {P.~G.}\ \bibnamefont
  {Wolynes}},\ }\href@noop {} {\bibfield  {journal} {\bibinfo  {journal}
  {Physical Review Letters}\ }\textbf {\bibinfo {volume} {86}},\ \bibinfo
  {pages} {5526} (\bibinfo {year} {2001})}\BibitemShut {NoStop}%
\bibitem [{\citenamefont {Stevenson}\ \emph {et~al.}(2008)\citenamefont
  {Stevenson}, \citenamefont {Walczak}, \citenamefont {Hall},\ and\
  \citenamefont {Wolynes}}]{doi:10.1063/1.3009827}%
  \BibitemOpen
  \bibfield  {author} {\bibinfo {author} {\bibfnamefont {J.~D.}\ \bibnamefont
  {Stevenson}}, \bibinfo {author} {\bibfnamefont {A.~M.}\ \bibnamefont
  {Walczak}}, \bibinfo {author} {\bibfnamefont {R.~W.}\ \bibnamefont {Hall}}, \
  and\ \bibinfo {author} {\bibfnamefont {P.~G.}\ \bibnamefont {Wolynes}},\
  }\href {\doibase 10.1063/1.3009827} {\bibfield  {journal} {\bibinfo
  {journal} {The Journal of Chemical Physics}\ }\textbf {\bibinfo {volume}
  {129}},\ \bibinfo {pages} {194505} (\bibinfo {year} {2008})}\BibitemShut
  {NoStop}%
\bibitem [{\citenamefont {Biroli}\ \emph
  {et~al.}(2014{\natexlab{a}})\citenamefont {Biroli}, \citenamefont
  {Cammarota}, \citenamefont {Tarjus},\ and\ \citenamefont
  {Tarzia}}]{PhysRevLett.112.175701}%
  \BibitemOpen
  \bibfield  {author} {\bibinfo {author} {\bibfnamefont {G.}~\bibnamefont
  {Biroli}}, \bibinfo {author} {\bibfnamefont {C.}~\bibnamefont {Cammarota}},
  \bibinfo {author} {\bibfnamefont {G.}~\bibnamefont {Tarjus}}, \ and\ \bibinfo
  {author} {\bibfnamefont {M.}~\bibnamefont {Tarzia}},\ }\href {\doibase
  10.1103/PhysRevLett.112.175701} {\bibfield  {journal} {\bibinfo  {journal}
  {Phys. Rev. Lett.}\ }\textbf {\bibinfo {volume} {112}},\ \bibinfo {pages}
  {175701} (\bibinfo {year} {2014}{\natexlab{a}})}\BibitemShut {NoStop}%
\bibitem [{\citenamefont {Dzero}\ \emph {et~al.}(2009)\citenamefont {Dzero},
  \citenamefont {Schmalian},\ and\ \citenamefont {Wolynes}}]{dzero2009replica}%
  \BibitemOpen
  \bibfield  {author} {\bibinfo {author} {\bibfnamefont {M.}~\bibnamefont
  {Dzero}}, \bibinfo {author} {\bibfnamefont {J.}~\bibnamefont {Schmalian}}, \
  and\ \bibinfo {author} {\bibfnamefont {P.~G.}\ \bibnamefont {Wolynes}},\
  }\href@noop {} {\bibfield  {journal} {\bibinfo  {journal} {Physical Review
  B}\ }\textbf {\bibinfo {volume} {80}},\ \bibinfo {pages} {024204} (\bibinfo
  {year} {2009})}\BibitemShut {NoStop}%
\bibitem [{\citenamefont {Böhmer}\ \emph {et~al.}(1993)\citenamefont
  {Böhmer}, \citenamefont {Ngai}, \citenamefont {Angell},\ and\ \citenamefont
  {Plazek}}]{doi:10.1063/1.466117}%
  \BibitemOpen
  \bibfield  {author} {\bibinfo {author} {\bibfnamefont {R.}~\bibnamefont
  {Böhmer}}, \bibinfo {author} {\bibfnamefont {K.~L.}\ \bibnamefont {Ngai}},
  \bibinfo {author} {\bibfnamefont {C.~A.}\ \bibnamefont {Angell}}, \ and\
  \bibinfo {author} {\bibfnamefont {D.~J.}\ \bibnamefont {Plazek}},\ }\href
  {\doibase 10.1063/1.466117} {\bibfield  {journal} {\bibinfo  {journal} {The
  Journal of Chemical Physics}\ }\textbf {\bibinfo {volume} {99}},\ \bibinfo
  {pages} {4201} (\bibinfo {year} {1993})}\BibitemShut {NoStop}%
\bibitem [{\citenamefont {Cardona}\ \emph {et~al.}(2007)\citenamefont
  {Cardona}, \citenamefont {Chamberlin},\ and\ \citenamefont
  {Marx}}]{cardona2007history}%
  \BibitemOpen
  \bibfield  {author} {\bibinfo {author} {\bibfnamefont {M.}~\bibnamefont
  {Cardona}}, \bibinfo {author} {\bibfnamefont {R.}~\bibnamefont {Chamberlin}},
  \ and\ \bibinfo {author} {\bibfnamefont {W.}~\bibnamefont {Marx}},\
  }\href@noop {} {\bibfield  {journal} {\bibinfo  {journal} {Annalen der
  Physik}\ }\textbf {\bibinfo {volume} {16}},\ \bibinfo {pages} {842} (\bibinfo
  {year} {2007})}\BibitemShut {NoStop}%
\bibitem [{\citenamefont {Zondervan}\ \emph {et~al.}(2007)\citenamefont
  {Zondervan}, \citenamefont {Kulzer}, \citenamefont {Berkhout},\ and\
  \citenamefont {Orrit}}]{zondervan2007local}%
  \BibitemOpen
  \bibfield  {author} {\bibinfo {author} {\bibfnamefont {R.}~\bibnamefont
  {Zondervan}}, \bibinfo {author} {\bibfnamefont {F.}~\bibnamefont {Kulzer}},
  \bibinfo {author} {\bibfnamefont {G.~C.}\ \bibnamefont {Berkhout}}, \ and\
  \bibinfo {author} {\bibfnamefont {M.}~\bibnamefont {Orrit}},\ }\href@noop {}
  {\bibfield  {journal} {\bibinfo  {journal} {Proceedings of the National
  Academy of Sciences}\ }\textbf {\bibinfo {volume} {104}},\ \bibinfo {pages}
  {12628} (\bibinfo {year} {2007})}\BibitemShut {NoStop}%
\bibitem [{\citenamefont {Paeng}\ \emph {et~al.}(2015)\citenamefont {Paeng},
  \citenamefont {Park}, \citenamefont {Hoang},\ and\ \citenamefont
  {Kaufman}}]{paeng2015ideal}%
  \BibitemOpen
  \bibfield  {author} {\bibinfo {author} {\bibfnamefont {K.}~\bibnamefont
  {Paeng}}, \bibinfo {author} {\bibfnamefont {H.}~\bibnamefont {Park}},
  \bibinfo {author} {\bibfnamefont {D.~T.}\ \bibnamefont {Hoang}}, \ and\
  \bibinfo {author} {\bibfnamefont {L.~J.}\ \bibnamefont {Kaufman}},\
  }\href@noop {} {\bibfield  {journal} {\bibinfo  {journal} {Proceedings of the
  National Academy of Sciences}\ }\textbf {\bibinfo {volume} {112}},\ \bibinfo
  {pages} {4952} (\bibinfo {year} {2015})}\BibitemShut {NoStop}%
\bibitem [{\citenamefont {Manz}\ \emph {et~al.}(2018)\citenamefont {Manz},
  \citenamefont {Paeng},\ and\ \citenamefont {Kaufman}}]{manz2018single}%
  \BibitemOpen
  \bibfield  {author} {\bibinfo {author} {\bibfnamefont {A.~S.}\ \bibnamefont
  {Manz}}, \bibinfo {author} {\bibfnamefont {K.}~\bibnamefont {Paeng}}, \ and\
  \bibinfo {author} {\bibfnamefont {L.~J.}\ \bibnamefont {Kaufman}},\
  }\href@noop {} {\bibfield  {journal} {\bibinfo  {journal} {The Journal of
  Chemical Physics}\ }\textbf {\bibinfo {volume} {148}},\ \bibinfo {pages}
  {204508} (\bibinfo {year} {2018})}\BibitemShut {NoStop}%
\bibitem [{\citenamefont {Manz}\ \emph {et~al.}(2019)\citenamefont {Manz},
  \citenamefont {Aly},\ and\ \citenamefont {Kaufman}}]{manz2019correlating}%
  \BibitemOpen
  \bibfield  {author} {\bibinfo {author} {\bibfnamefont {A.~S.}\ \bibnamefont
  {Manz}}, \bibinfo {author} {\bibfnamefont {M.}~\bibnamefont {Aly}}, \ and\
  \bibinfo {author} {\bibfnamefont {L.~J.}\ \bibnamefont {Kaufman}},\
  }\href@noop {} {\bibfield  {journal} {\bibinfo  {journal} {The Journal of
  chemical physics}\ }\textbf {\bibinfo {volume} {151}},\ \bibinfo {pages}
  {084501} (\bibinfo {year} {2019})}\BibitemShut {NoStop}%
\bibitem [{\citenamefont {Zhang}\ \emph {et~al.}(2018)\citenamefont {Zhang},
  \citenamefont {Maldonis}, \citenamefont {Liu}, \citenamefont {Schroers},\
  and\ \citenamefont {Voyles}}]{zhang2018spatially}%
  \BibitemOpen
  \bibfield  {author} {\bibinfo {author} {\bibfnamefont {P.}~\bibnamefont
  {Zhang}}, \bibinfo {author} {\bibfnamefont {J.~J.}\ \bibnamefont {Maldonis}},
  \bibinfo {author} {\bibfnamefont {Z.}~\bibnamefont {Liu}}, \bibinfo {author}
  {\bibfnamefont {J.}~\bibnamefont {Schroers}}, \ and\ \bibinfo {author}
  {\bibfnamefont {P.~M.}\ \bibnamefont {Voyles}},\ }\href@noop {} {\bibfield
  {journal} {\bibinfo  {journal} {Nature communications}\ }\textbf {\bibinfo
  {volume} {9}},\ \bibinfo {pages} {1} (\bibinfo {year} {2018})}\BibitemShut
  {NoStop}%
\bibitem [{\citenamefont {Shang}\ \emph {et~al.}(2019)\citenamefont {Shang},
  \citenamefont {Rottler}, \citenamefont {Guan},\ and\ \citenamefont
  {Barrat}}]{shang2019local}%
  \BibitemOpen
  \bibfield  {author} {\bibinfo {author} {\bibfnamefont {B.}~\bibnamefont
  {Shang}}, \bibinfo {author} {\bibfnamefont {J.}~\bibnamefont {Rottler}},
  \bibinfo {author} {\bibfnamefont {P.}~\bibnamefont {Guan}}, \ and\ \bibinfo
  {author} {\bibfnamefont {J.-L.}\ \bibnamefont {Barrat}},\ }\href@noop {}
  {\bibfield  {journal} {\bibinfo  {journal} {Physical review letters}\
  }\textbf {\bibinfo {volume} {122}},\ \bibinfo {pages} {105501} (\bibinfo
  {year} {2019})}\BibitemShut {NoStop}%
\bibitem [{\citenamefont {Widmer-Cooper}\ \emph {et~al.}(2008)\citenamefont
  {Widmer-Cooper}, \citenamefont {Perry}, \citenamefont {Harrowell},\ and\
  \citenamefont {Reichman}}]{widmer2008irreversible}%
  \BibitemOpen
  \bibfield  {author} {\bibinfo {author} {\bibfnamefont {A.}~\bibnamefont
  {Widmer-Cooper}}, \bibinfo {author} {\bibfnamefont {H.}~\bibnamefont
  {Perry}}, \bibinfo {author} {\bibfnamefont {P.}~\bibnamefont {Harrowell}}, \
  and\ \bibinfo {author} {\bibfnamefont {D.~R.}\ \bibnamefont {Reichman}},\
  }\href@noop {} {\bibfield  {journal} {\bibinfo  {journal} {Nature Physics}\
  }\textbf {\bibinfo {volume} {4}},\ \bibinfo {pages} {711} (\bibinfo {year}
  {2008})}\BibitemShut {NoStop}%
\bibitem [{\citenamefont {Schoenholz}\ \emph {et~al.}(2016)\citenamefont
  {Schoenholz}, \citenamefont {Cubuk}, \citenamefont {Sussman}, \citenamefont
  {Kaxiras},\ and\ \citenamefont {Liu}}]{schoenholz2016structural}%
  \BibitemOpen
  \bibfield  {author} {\bibinfo {author} {\bibfnamefont {S.~S.}\ \bibnamefont
  {Schoenholz}}, \bibinfo {author} {\bibfnamefont {E.~D.}\ \bibnamefont
  {Cubuk}}, \bibinfo {author} {\bibfnamefont {D.~M.}\ \bibnamefont {Sussman}},
  \bibinfo {author} {\bibfnamefont {E.}~\bibnamefont {Kaxiras}}, \ and\
  \bibinfo {author} {\bibfnamefont {A.~J.}\ \bibnamefont {Liu}},\ }\href@noop
  {} {\bibfield  {journal} {\bibinfo  {journal} {Nature Physics}\ }\textbf
  {\bibinfo {volume} {12}},\ \bibinfo {pages} {469} (\bibinfo {year}
  {2016})}\BibitemShut {NoStop}%
\bibitem [{\citenamefont {Tong}\ and\ \citenamefont
  {Tanaka}(2018)}]{tong2018revealing}%
  \BibitemOpen
  \bibfield  {author} {\bibinfo {author} {\bibfnamefont {H.}~\bibnamefont
  {Tong}}\ and\ \bibinfo {author} {\bibfnamefont {H.}~\bibnamefont {Tanaka}},\
  }\href@noop {} {\bibfield  {journal} {\bibinfo  {journal} {Physical Review
  X}\ }\textbf {\bibinfo {volume} {8}},\ \bibinfo {pages} {011041} (\bibinfo
  {year} {2018})}\BibitemShut {NoStop}%
\bibitem [{\citenamefont {Richert}\ and\ \citenamefont
  {Angell}(1998)}]{richert1998dynamics}%
  \BibitemOpen
  \bibfield  {author} {\bibinfo {author} {\bibfnamefont {R.}~\bibnamefont
  {Richert}}\ and\ \bibinfo {author} {\bibfnamefont {C.}~\bibnamefont
  {Angell}},\ }\href@noop {} {\bibfield  {journal} {\bibinfo  {journal} {J.
  Chem. Phys.}\ }\textbf {\bibinfo {volume} {108}},\ \bibinfo {pages} {9016}
  (\bibinfo {year} {1998})}\BibitemShut {NoStop}%
\bibitem [{\citenamefont {Sastry}(2001)}]{sastry2001relationship}%
  \BibitemOpen
  \bibfield  {author} {\bibinfo {author} {\bibfnamefont {S.}~\bibnamefont
  {Sastry}},\ }\href@noop {} {\bibfield  {journal} {\bibinfo  {journal}
  {Nature}\ }\textbf {\bibinfo {volume} {409}},\ \bibinfo {pages} {164}
  (\bibinfo {year} {2001})}\BibitemShut {NoStop}%
\bibitem [{\citenamefont {Larini}\ \emph {et~al.}(2008)\citenamefont {Larini},
  \citenamefont {Ottochian}, \citenamefont {De~Michele},\ and\ \citenamefont
  {Leporini}}]{larini2008universal}%
  \BibitemOpen
  \bibfield  {author} {\bibinfo {author} {\bibfnamefont {L.}~\bibnamefont
  {Larini}}, \bibinfo {author} {\bibfnamefont {A.}~\bibnamefont {Ottochian}},
  \bibinfo {author} {\bibfnamefont {C.}~\bibnamefont {De~Michele}}, \ and\
  \bibinfo {author} {\bibfnamefont {D.}~\bibnamefont {Leporini}},\ }\href@noop
  {} {\bibfield  {journal} {\bibinfo  {journal} {Nature Physics}\ }\textbf
  {\bibinfo {volume} {4}},\ \bibinfo {pages} {42} (\bibinfo {year}
  {2008})}\BibitemShut {NoStop}%
\bibitem [{\citenamefont {Gundermann}\ \emph {et~al.}(2014)\citenamefont
  {Gundermann}, \citenamefont {Niss}, \citenamefont {Christensen},
  \citenamefont {Dyre},\ and\ \citenamefont
  {Hecksher}}]{gundermann2014dynamic}%
  \BibitemOpen
  \bibfield  {author} {\bibinfo {author} {\bibfnamefont {D.}~\bibnamefont
  {Gundermann}}, \bibinfo {author} {\bibfnamefont {K.}~\bibnamefont {Niss}},
  \bibinfo {author} {\bibfnamefont {T.}~\bibnamefont {Christensen}}, \bibinfo
  {author} {\bibfnamefont {J.~C.}\ \bibnamefont {Dyre}}, \ and\ \bibinfo
  {author} {\bibfnamefont {T.}~\bibnamefont {Hecksher}},\ }\href@noop {}
  {\bibfield  {journal} {\bibinfo  {journal} {The Journal of chemical physics}\
  }\textbf {\bibinfo {volume} {140}},\ \bibinfo {pages} {244508} (\bibinfo
  {year} {2014})}\BibitemShut {NoStop}%
\bibitem [{\citenamefont {Ozawa}\ \emph {et~al.}(2019)\citenamefont {Ozawa},
  \citenamefont {Scalliet}, \citenamefont {Ninarello},\ and\ \citenamefont
  {Berthier}}]{ozawa2019does}%
  \BibitemOpen
  \bibfield  {author} {\bibinfo {author} {\bibfnamefont {M.}~\bibnamefont
  {Ozawa}}, \bibinfo {author} {\bibfnamefont {C.}~\bibnamefont {Scalliet}},
  \bibinfo {author} {\bibfnamefont {A.}~\bibnamefont {Ninarello}}, \ and\
  \bibinfo {author} {\bibfnamefont {L.}~\bibnamefont {Berthier}},\ }\href@noop
  {} {\bibfield  {journal} {\bibinfo  {journal} {J. Chem. Phys.}\ }\textbf
  {\bibinfo {volume} {151}},\ \bibinfo {pages} {084504} (\bibinfo {year}
  {2019})}\BibitemShut {NoStop}%
\bibitem [{\citenamefont {Kirkpatrick}\ \emph {et~al.}(1989)\citenamefont
  {Kirkpatrick}, \citenamefont {Thirumalai},\ and\ \citenamefont
  {Wolynes}}]{kirkpatrick1989scaling}%
  \BibitemOpen
  \bibfield  {author} {\bibinfo {author} {\bibfnamefont {T.~R.}\ \bibnamefont
  {Kirkpatrick}}, \bibinfo {author} {\bibfnamefont {D.}~\bibnamefont
  {Thirumalai}}, \ and\ \bibinfo {author} {\bibfnamefont {P.~G.}\ \bibnamefont
  {Wolynes}},\ }\href@noop {} {\bibfield  {journal} {\bibinfo  {journal} {Phys.
  Rev. A}\ }\textbf {\bibinfo {volume} {40}},\ \bibinfo {pages} {1045}
  (\bibinfo {year} {1989})}\BibitemShut {NoStop}%
\bibitem [{\citenamefont {Berthier}\ \emph {et~al.}(2019)\citenamefont
  {Berthier}, \citenamefont {Ozawa},\ and\ \citenamefont
  {Scalliet}}]{berthier2019configurational}%
  \BibitemOpen
  \bibfield  {author} {\bibinfo {author} {\bibfnamefont {L.}~\bibnamefont
  {Berthier}}, \bibinfo {author} {\bibfnamefont {M.}~\bibnamefont {Ozawa}}, \
  and\ \bibinfo {author} {\bibfnamefont {C.}~\bibnamefont {Scalliet}},\
  }\href@noop {} {\bibfield  {journal} {\bibinfo  {journal} {J. Chem. Phys.}\
  }\textbf {\bibinfo {volume} {150}},\ \bibinfo {pages} {160902} (\bibinfo
  {year} {2019})}\BibitemShut {NoStop}%
\bibitem [{\citenamefont {Berthier}\ and\ \citenamefont
  {Jack}(2015)}]{berthier2015evidence}%
  \BibitemOpen
  \bibfield  {author} {\bibinfo {author} {\bibfnamefont {L.}~\bibnamefont
  {Berthier}}\ and\ \bibinfo {author} {\bibfnamefont {R.~L.}\ \bibnamefont
  {Jack}},\ }\href@noop {} {\bibfield  {journal} {\bibinfo  {journal} {Physical
  review letters}\ }\textbf {\bibinfo {volume} {114}},\ \bibinfo {pages}
  {205701} (\bibinfo {year} {2015})}\BibitemShut {NoStop}%
\bibitem [{\citenamefont {Coslovich}\ and\ \citenamefont
  {Jack}(2016)}]{coslovich2016structure}%
  \BibitemOpen
  \bibfield  {author} {\bibinfo {author} {\bibfnamefont {D.}~\bibnamefont
  {Coslovich}}\ and\ \bibinfo {author} {\bibfnamefont {R.~L.}\ \bibnamefont
  {Jack}},\ }\href@noop {} {\bibfield  {journal} {\bibinfo  {journal} {Journal
  of Statistical Mechanics: Theory and Experiment}\ }\textbf {\bibinfo {volume}
  {2016}},\ \bibinfo {pages} {074012} (\bibinfo {year} {2016})}\BibitemShut
  {NoStop}%
\bibitem [{\citenamefont {Guiselin}\ \emph {et~al.}(2020)\citenamefont
  {Guiselin}, \citenamefont {Berthier},\ and\ \citenamefont
  {Tarjus}}]{guiselin2020random}%
  \BibitemOpen
  \bibfield  {author} {\bibinfo {author} {\bibfnamefont {B.}~\bibnamefont
  {Guiselin}}, \bibinfo {author} {\bibfnamefont {L.}~\bibnamefont {Berthier}},
  \ and\ \bibinfo {author} {\bibfnamefont {G.}~\bibnamefont {Tarjus}},\
  }\href@noop {} {\bibfield  {journal} {\bibinfo  {journal} {Physical Review
  E}\ }\textbf {\bibinfo {volume} {102}},\ \bibinfo {pages} {042129} (\bibinfo
  {year} {2020})}\BibitemShut {NoStop}%
\bibitem [{\citenamefont {Adam}\ and\ \citenamefont
  {Gibbs}(1965)}]{adam1965temperature}%
  \BibitemOpen
  \bibfield  {author} {\bibinfo {author} {\bibfnamefont {G.}~\bibnamefont
  {Adam}}\ and\ \bibinfo {author} {\bibfnamefont {J.~H.}\ \bibnamefont
  {Gibbs}},\ }\href@noop {} {\bibfield  {journal} {\bibinfo  {journal} {J.
  Chem. Phys.}\ }\textbf {\bibinfo {volume} {43}},\ \bibinfo {pages} {139}
  (\bibinfo {year} {1965})}\BibitemShut {NoStop}%
\bibitem [{\citenamefont {B{\"u}chner}\ and\ \citenamefont
  {Heuer}(1999)}]{buchner1999potential}%
  \BibitemOpen
  \bibfield  {author} {\bibinfo {author} {\bibfnamefont {S.}~\bibnamefont
  {B{\"u}chner}}\ and\ \bibinfo {author} {\bibfnamefont {A.}~\bibnamefont
  {Heuer}},\ }\href@noop {} {\bibfield  {journal} {\bibinfo  {journal}
  {Physical Review E}\ }\textbf {\bibinfo {volume} {60}},\ \bibinfo {pages}
  {6507} (\bibinfo {year} {1999})}\BibitemShut {NoStop}%
\bibitem [{\citenamefont {La~Nave}\ \emph {et~al.}(2006)\citenamefont
  {La~Nave}, \citenamefont {Sastry},\ and\ \citenamefont
  {Sciortino}}]{la2006relation}%
  \BibitemOpen
  \bibfield  {author} {\bibinfo {author} {\bibfnamefont {E.}~\bibnamefont
  {La~Nave}}, \bibinfo {author} {\bibfnamefont {S.}~\bibnamefont {Sastry}}, \
  and\ \bibinfo {author} {\bibfnamefont {F.}~\bibnamefont {Sciortino}},\
  }\href@noop {} {\bibfield  {journal} {\bibinfo  {journal} {Physical Review
  E}\ }\textbf {\bibinfo {volume} {74}},\ \bibinfo {pages} {050501} (\bibinfo
  {year} {2006})}\BibitemShut {NoStop}%
\bibitem [{\citenamefont {Heuer}(2008)}]{heuer2008exploring}%
  \BibitemOpen
  \bibfield  {author} {\bibinfo {author} {\bibfnamefont {A.}~\bibnamefont
  {Heuer}},\ }\href@noop {} {\bibfield  {journal} {\bibinfo  {journal} {Journal
  of Physics: Condensed Matter}\ }\textbf {\bibinfo {volume} {20}},\ \bibinfo
  {pages} {373101} (\bibinfo {year} {2008})}\BibitemShut {NoStop}%
\bibitem [{\citenamefont {Ninarello}\ \emph {et~al.}(2017)\citenamefont
  {Ninarello}, \citenamefont {Berthier},\ and\ \citenamefont
  {Coslovich}}]{ninarello2017models}%
  \BibitemOpen
  \bibfield  {author} {\bibinfo {author} {\bibfnamefont {A.}~\bibnamefont
  {Ninarello}}, \bibinfo {author} {\bibfnamefont {L.}~\bibnamefont {Berthier}},
  \ and\ \bibinfo {author} {\bibfnamefont {D.}~\bibnamefont {Coslovich}},\
  }\href@noop {} {\bibfield  {journal} {\bibinfo  {journal} {Phys. Rev. X}\
  }\textbf {\bibinfo {volume} {7}},\ \bibinfo {pages} {021039} (\bibinfo {year}
  {2017})}\BibitemShut {NoStop}%
\bibitem [{\citenamefont {Widmer-Cooper}\ \emph {et~al.}(2004)\citenamefont
  {Widmer-Cooper}, \citenamefont {Harrowell},\ and\ \citenamefont
  {Fynewever}}]{widmer2004reproducible}%
  \BibitemOpen
  \bibfield  {author} {\bibinfo {author} {\bibfnamefont {A.}~\bibnamefont
  {Widmer-Cooper}}, \bibinfo {author} {\bibfnamefont {P.}~\bibnamefont
  {Harrowell}}, \ and\ \bibinfo {author} {\bibfnamefont {H.}~\bibnamefont
  {Fynewever}},\ }\href@noop {} {\bibfield  {journal} {\bibinfo  {journal}
  {Physical review letters}\ }\textbf {\bibinfo {volume} {93}},\ \bibinfo
  {pages} {135701} (\bibinfo {year} {2004})}\BibitemShut {NoStop}%
\bibitem [{\citenamefont {Franz}\ and\ \citenamefont
  {Parisi}(1997)}]{franz1997phase}%
  \BibitemOpen
  \bibfield  {author} {\bibinfo {author} {\bibfnamefont {S.}~\bibnamefont
  {Franz}}\ and\ \bibinfo {author} {\bibfnamefont {G.}~\bibnamefont {Parisi}},\
  }\href@noop {} {\bibfield  {journal} {\bibinfo  {journal} {Physical review
  letters}\ }\textbf {\bibinfo {volume} {79}},\ \bibinfo {pages} {2486}
  (\bibinfo {year} {1997})}\BibitemShut {NoStop}%
\bibitem [{\citenamefont {Berthier}\ and\ \citenamefont
  {Coslovich}(2014)}]{berthier2014novel}%
  \BibitemOpen
  \bibfield  {author} {\bibinfo {author} {\bibfnamefont {L.}~\bibnamefont
  {Berthier}}\ and\ \bibinfo {author} {\bibfnamefont {D.}~\bibnamefont
  {Coslovich}},\ }\href@noop {} {\bibfield  {journal} {\bibinfo  {journal}
  {PNAS}\ }\textbf {\bibinfo {volume} {111}},\ \bibinfo {pages} {11668}
  (\bibinfo {year} {2014})}\BibitemShut {NoStop}%
\bibitem [{\citenamefont {Berthier}\ and\ \citenamefont
  {Ediger}(2020)}]{berthier2020measure}%
  \BibitemOpen
  \bibfield  {author} {\bibinfo {author} {\bibfnamefont {L.}~\bibnamefont
  {Berthier}}\ and\ \bibinfo {author} {\bibfnamefont {M.~D.}\ \bibnamefont
  {Ediger}},\ }\href@noop {} {\bibfield  {journal} {\bibinfo  {journal} {The
  Journal of Chemical Physics}\ }\textbf {\bibinfo {volume} {153}},\ \bibinfo
  {pages} {044501} (\bibinfo {year} {2020})}\BibitemShut {NoStop}%
\bibitem [{\citenamefont {Brumer}\ and\ \citenamefont
  {Reichman}(2004)}]{brumer2004numerical}%
  \BibitemOpen
  \bibfield  {author} {\bibinfo {author} {\bibfnamefont {Y.}~\bibnamefont
  {Brumer}}\ and\ \bibinfo {author} {\bibfnamefont {D.~R.}\ \bibnamefont
  {Reichman}},\ }\href@noop {} {\bibfield  {journal} {\bibinfo  {journal} {The
  Journal of Physical Chemistry B}\ }\textbf {\bibinfo {volume} {108}},\
  \bibinfo {pages} {6832} (\bibinfo {year} {2004})}\BibitemShut {NoStop}%
\bibitem [{\citenamefont {Berthier}\ and\ \citenamefont
  {Witten}(2009)}]{berthier2009glass}%
  \BibitemOpen
  \bibfield  {author} {\bibinfo {author} {\bibfnamefont {L.}~\bibnamefont
  {Berthier}}\ and\ \bibinfo {author} {\bibfnamefont {T.~A.}\ \bibnamefont
  {Witten}},\ }\href@noop {} {\bibfield  {journal} {\bibinfo  {journal} {Phys.
  Rev. E}\ }\textbf {\bibinfo {volume} {80}},\ \bibinfo {pages} {021502}
  (\bibinfo {year} {2009})}\BibitemShut {NoStop}%
\bibitem [{\citenamefont {Berthier}\ \emph {et~al.}(2017)\citenamefont
  {Berthier}, \citenamefont {Charbonneau}, \citenamefont {Coslovich},
  \citenamefont {Ninarello}, \citenamefont {Ozawa},\ and\ \citenamefont
  {Yaida}}]{berthier2017configurational}%
  \BibitemOpen
  \bibfield  {author} {\bibinfo {author} {\bibfnamefont {L.}~\bibnamefont
  {Berthier}}, \bibinfo {author} {\bibfnamefont {P.}~\bibnamefont
  {Charbonneau}}, \bibinfo {author} {\bibfnamefont {D.}~\bibnamefont
  {Coslovich}}, \bibinfo {author} {\bibfnamefont {A.}~\bibnamefont
  {Ninarello}}, \bibinfo {author} {\bibfnamefont {M.}~\bibnamefont {Ozawa}}, \
  and\ \bibinfo {author} {\bibfnamefont {S.}~\bibnamefont {Yaida}},\
  }\href@noop {} {\bibfield  {journal} {\bibinfo  {journal} {PNAS}\ }\textbf
  {\bibinfo {volume} {114}},\ \bibinfo {pages} {11356} (\bibinfo {year}
  {2017})}\BibitemShut {NoStop}%
\bibitem [{\citenamefont {Ozawa}\ and\ \citenamefont
  {Berthier}(2017)}]{ozawa2017does}%
  \BibitemOpen
  \bibfield  {author} {\bibinfo {author} {\bibfnamefont {M.}~\bibnamefont
  {Ozawa}}\ and\ \bibinfo {author} {\bibfnamefont {L.}~\bibnamefont
  {Berthier}},\ }\href@noop {} {\bibfield  {journal} {\bibinfo  {journal} {J.
  Chem. Phys.}\ }\textbf {\bibinfo {volume} {146}},\ \bibinfo {pages} {014502}
  (\bibinfo {year} {2017})}\BibitemShut {NoStop}%
\bibitem [{\citenamefont {B{\"o}hmer}\ \emph {et~al.}(1998)\citenamefont
  {B{\"o}hmer}, \citenamefont {Chamberlin}, \citenamefont {Diezemann},
  \citenamefont {Geil}, \citenamefont {Heuer}, \citenamefont {Hinze},
  \citenamefont {Kuebler}, \citenamefont {Richert}, \citenamefont {Schiener},
  \citenamefont {Sillescu} \emph {et~al.}}]{bohmer1998nature}%
  \BibitemOpen
  \bibfield  {author} {\bibinfo {author} {\bibfnamefont {R.}~\bibnamefont
  {B{\"o}hmer}}, \bibinfo {author} {\bibfnamefont {R.}~\bibnamefont
  {Chamberlin}}, \bibinfo {author} {\bibfnamefont {G.}~\bibnamefont
  {Diezemann}}, \bibinfo {author} {\bibfnamefont {B.}~\bibnamefont {Geil}},
  \bibinfo {author} {\bibfnamefont {A.}~\bibnamefont {Heuer}}, \bibinfo
  {author} {\bibfnamefont {G.}~\bibnamefont {Hinze}}, \bibinfo {author}
  {\bibfnamefont {S.}~\bibnamefont {Kuebler}}, \bibinfo {author} {\bibfnamefont
  {R.}~\bibnamefont {Richert}}, \bibinfo {author} {\bibfnamefont
  {B.}~\bibnamefont {Schiener}}, \bibinfo {author} {\bibfnamefont
  {H.}~\bibnamefont {Sillescu}},  \emph {et~al.},\ }\href@noop {} {\bibfield
  {journal} {\bibinfo  {journal} {Journal of non-crystalline solids}\ }\textbf
  {\bibinfo {volume} {235}},\ \bibinfo {pages} {1} (\bibinfo {year}
  {1998})}\BibitemShut {NoStop}%
\bibitem [{\citenamefont {Angell}\ \emph {et~al.}(2000)\citenamefont {Angell},
  \citenamefont {Ngai}, \citenamefont {McKenna}, \citenamefont {McMillan},\
  and\ \citenamefont {Martin}}]{angell2000relaxation}%
  \BibitemOpen
  \bibfield  {author} {\bibinfo {author} {\bibfnamefont {C.~A.}\ \bibnamefont
  {Angell}}, \bibinfo {author} {\bibfnamefont {K.~L.}\ \bibnamefont {Ngai}},
  \bibinfo {author} {\bibfnamefont {G.~B.}\ \bibnamefont {McKenna}}, \bibinfo
  {author} {\bibfnamefont {P.~F.}\ \bibnamefont {McMillan}}, \ and\ \bibinfo
  {author} {\bibfnamefont {S.~W.}\ \bibnamefont {Martin}},\ }\href@noop {}
  {\bibfield  {journal} {\bibinfo  {journal} {Journal of applied physics}\
  }\textbf {\bibinfo {volume} {88}},\ \bibinfo {pages} {3113} (\bibinfo {year}
  {2000})}\BibitemShut {NoStop}%
\bibitem [{\citenamefont {Kuebler}\ \emph {et~al.}(1997)\citenamefont
  {Kuebler}, \citenamefont {Heuer},\ and\ \citenamefont
  {Spiess}}]{kuebler1997glass}%
  \BibitemOpen
  \bibfield  {author} {\bibinfo {author} {\bibfnamefont {S.}~\bibnamefont
  {Kuebler}}, \bibinfo {author} {\bibfnamefont {A.}~\bibnamefont {Heuer}}, \
  and\ \bibinfo {author} {\bibfnamefont {H.~W.}\ \bibnamefont {Spiess}},\
  }\href@noop {} {\bibfield  {journal} {\bibinfo  {journal} {Physical Review
  E}\ }\textbf {\bibinfo {volume} {56}},\ \bibinfo {pages} {741} (\bibinfo
  {year} {1997})}\BibitemShut {NoStop}%
\bibitem [{\citenamefont {Berthier}(2013)}]{berthier2013overlap}%
  \BibitemOpen
  \bibfield  {author} {\bibinfo {author} {\bibfnamefont {L.}~\bibnamefont
  {Berthier}},\ }\href@noop {} {\bibfield  {journal} {\bibinfo  {journal}
  {Physical Review E}\ }\textbf {\bibinfo {volume} {88}},\ \bibinfo {pages}
  {022313} (\bibinfo {year} {2013})}\BibitemShut {NoStop}%
\bibitem [{\citenamefont {Biroli}\ \emph
  {et~al.}(2014{\natexlab{b}})\citenamefont {Biroli}, \citenamefont
  {Cammarota}, \citenamefont {Tarjus},\ and\ \citenamefont
  {Tarzia}}]{biroli2014random}%
  \BibitemOpen
  \bibfield  {author} {\bibinfo {author} {\bibfnamefont {G.}~\bibnamefont
  {Biroli}}, \bibinfo {author} {\bibfnamefont {C.}~\bibnamefont {Cammarota}},
  \bibinfo {author} {\bibfnamefont {G.}~\bibnamefont {Tarjus}}, \ and\ \bibinfo
  {author} {\bibfnamefont {M.}~\bibnamefont {Tarzia}},\ }\href@noop {}
  {\bibfield  {journal} {\bibinfo  {journal} {Physical review letters}\
  }\textbf {\bibinfo {volume} {112}},\ \bibinfo {pages} {175701} (\bibinfo
  {year} {2014}{\natexlab{b}})}\BibitemShut {NoStop}%
\bibitem [{\citenamefont {Biroli}\ \emph {et~al.}(2018)\citenamefont {Biroli},
  \citenamefont {Cammarota}, \citenamefont {Tarjus},\ and\ \citenamefont
  {Tarzia}}]{biroli2018random}%
  \BibitemOpen
  \bibfield  {author} {\bibinfo {author} {\bibfnamefont {G.}~\bibnamefont
  {Biroli}}, \bibinfo {author} {\bibfnamefont {C.}~\bibnamefont {Cammarota}},
  \bibinfo {author} {\bibfnamefont {G.}~\bibnamefont {Tarjus}}, \ and\ \bibinfo
  {author} {\bibfnamefont {M.}~\bibnamefont {Tarzia}},\ }\href@noop {}
  {\bibfield  {journal} {\bibinfo  {journal} {Physical Review B}\ }\textbf
  {\bibinfo {volume} {98}},\ \bibinfo {pages} {174205} (\bibinfo {year}
  {2018})}\BibitemShut {NoStop}%
\bibitem [{\citenamefont {Berthier}\ and\ \citenamefont
  {Jack}(2007)}]{berthier2007structure}%
  \BibitemOpen
  \bibfield  {author} {\bibinfo {author} {\bibfnamefont {L.}~\bibnamefont
  {Berthier}}\ and\ \bibinfo {author} {\bibfnamefont {R.~L.}\ \bibnamefont
  {Jack}},\ }\href@noop {} {\bibfield  {journal} {\bibinfo  {journal} {Physical
  Review E}\ }\textbf {\bibinfo {volume} {76}},\ \bibinfo {pages} {041509}
  (\bibinfo {year} {2007})}\BibitemShut {NoStop}%
\bibitem [{\citenamefont {Jack}\ and\ \citenamefont
  {Garrahan}(2016)}]{jack2016phase}%
  \BibitemOpen
  \bibfield  {author} {\bibinfo {author} {\bibfnamefont {R.~L.}\ \bibnamefont
  {Jack}}\ and\ \bibinfo {author} {\bibfnamefont {J.~P.}\ \bibnamefont
  {Garrahan}},\ }\href@noop {} {\bibfield  {journal} {\bibinfo  {journal}
  {Physical review letters}\ }\textbf {\bibinfo {volume} {116}},\ \bibinfo
  {pages} {055702} (\bibinfo {year} {2016})}\BibitemShut {NoStop}%
\bibitem [{\citenamefont {Biroli}\ and\ \citenamefont
  {Bouchaud}(2012)}]{biroli2012random}%
  \BibitemOpen
  \bibfield  {author} {\bibinfo {author} {\bibfnamefont {G.}~\bibnamefont
  {Biroli}}\ and\ \bibinfo {author} {\bibfnamefont {J.-P.}\ \bibnamefont
  {Bouchaud}},\ }\href@noop {} {\bibfield  {journal} {\bibinfo  {journal}
  {Structural Glasses and Supercooled Liquids: Theory, Experiment, and
  Applications}\ ,\ \bibinfo {pages} {31}} (\bibinfo {year}
  {2012})}\BibitemShut {NoStop}%
\bibitem [{\citenamefont {Jack}\ and\ \citenamefont
  {Garrahan}(2005)}]{jack2005caging}%
  \BibitemOpen
  \bibfield  {author} {\bibinfo {author} {\bibfnamefont {R.~L.}\ \bibnamefont
  {Jack}}\ and\ \bibinfo {author} {\bibfnamefont {J.~P.}\ \bibnamefont
  {Garrahan}},\ }\href@noop {} {\bibfield  {journal} {\bibinfo  {journal} {The
  Journal of chemical physics}\ }\textbf {\bibinfo {volume} {123}},\ \bibinfo
  {pages} {164508} (\bibinfo {year} {2005})}\BibitemShut {NoStop}%
\end{thebibliography}%

\end{document}